\documentclass[hyper]{JHEP} 

\usepackage{epsfig}

\newcommand\fverb{\setbox\pippobox=\hbox\bgroup\verb}

\newcommand\fverbdo{\egroup\medskip\noindent%

            \fbox{\unhbox\pippobox}\ }

\newcommand\fverbit{\egroup\item[\fbox{\unhbox\pippobox}]}

\newbox\pippobox

\title{T-duality of Non-Relativistic String in Torsional 
Newton-Cartan Background}
\author{J. Kluso\v{n}\\
Department of
Theoretical Physics and Astrophysics\\
Faculty of Science, Masaryk University\\
Kotl\'{a}\v{r}sk\'{a} 2, 611 37, Brno\\
Czech Republic\\
E-mail: \email{klu@physics.muni.cz}} \preprint{}

 \abstract{In this short note we analyse T-duality properties
 of non-relativistic String in Torsional Newton-Cartan Background. We also determine
condition that ensures that non-relativistic string maps to 
non-relativistic string under T-duality.}

\def\tp{\tilde{p}}

\def\hG{\hat{G}}

\def\hB{\hat{B}}
\def\bA{\mathbf{A}}

\def\bW{\mathbf{W}}

\def\ty{\tilde{y}}

\def\tlambda{\tilde{\lambda}}
\def\bB{\mathbf{B}}

\def\tx{\tilde{x}}

\def\be{\begin{equation}}

\def\ee{\end{equation}}

\def\bea{\begin{eqnarray}}

\def\eea{\end{eqnarray}}

\def\bX{\mathbf{X}}
\def\bY{\mathbf{Y}}

\def\mH{\mathcal{H}}

\def\tG{\tilde{G}}

\def\bM{\mathbf{M}}

\newcommand{\hg}{\hat{g}}

\newcommand{\tX}{\tilde{X}}

\newcommand{\hh}{\hat{h}}

\def \bA{\mathbf{A}}

\newcommand{\mL}{\mathcal{L}}

\def\pb #1{\left\{#1\right\}}

\begin{document}
\section{Introduction and Summary}
Recently there was renewed interest in the Newton-Cartan geometry (NC)  and its torsional
generalization for the study of non-relativistic aspects of string theory and gravity, see for example \cite{Andringa:2010it}. It is interesting that there are currently two versions
of non-relativistic string theories. First one was proposed in
\cite{Harmark:2017rpg,Harmark:2018cdl} and corresponds to string on torsional NC geometry while the second one was firstly introduced in \cite{Andringa:2012uz}
\footnote{For related works, see 
\cite{Bergshoeff:2019pij,Kluson:2019uza,Kluson:2019ifd,Bergshoeff:2018vfn,Bergshoeff:2018yvt,Kluson:2018grx,Bergshoeff:2018vfn,Kluson:2018egd,Kluson:2018uss,Kluson:2017abm}.} and corresponds to string in stringy NC geometry. Stringy NC geometry is characterized by foliation of space-time into two longitudinal directions that, roughly speaking, correspond to world-sheet of fundamental string. 
 In \cite{Gomis:2019zyu}
beta function for string in stringy NC gravity was proposed 
which leads to the dynamical equations of non-relativistic stringy NC gravity. In case of string on torsional NC geometry beta function was calculated in 
\cite{Gallegos:2019icg} where again this beta function leads to dynamical equations
of motion of torsional NC gravity. This is very interesting consistency check which demonstrates that both non-relativistic string theories could be considered as UV completion of corresponding non-relativistic theories of gravity. Finally  it was shown in remarkable 
paper \cite{Harmark:2019upf} that these seemingly different string theories can 
be mapped into each other. 

In this work we will be interested in the first version of non-relativistic string theory that arises by null reduction of relativistic string. As was argued in our previous paper
\cite{Kluson:2019qgj}, following 
\cite{Bergshoeff:2018yvt}
T-duality along null dimension is rather subtle and needs careful treatment. In fact, it is convenient to consider extended string with two auxiliary fields and additional terms on its world-sheet so that the string now propagates  in the background with no null isometry. The meaning of two auxiliary fields is that
solving their equations of motion and plugging back to the action we get the original one. However since extended action has well defined kinetic term  it is much more convenient for performing T-duality as was shown in \cite{Kluson:2019qgj}. We performed it for general null metric and we found that the string in the background with null isometry is T-dual to non-relativistic string. 

In this work we apply this procedure to the concrete background \cite{Harmark:2019upf}. We explicitly derive corresponding Lagrangian density and we found that it has the same form as in \cite{Harmark:2017rpg}.
Then we address the main problem which is T-duality of non-relativistic string in torsional NC background. Since an action for non-relativistic string is non-linear in this background it is not clear whether it is possible to follow standard procedure that is based on the gauging 
isometry direction on the world-sheet of the string \cite{Buscher:1987sk,Buscher:1987qj}.
For that reason we mean that it is natural to start with extended relativistic background and perform T-duality along null direction together with T-duality along
one spatial direction. This problem is more complex and we were not able to solve it in the full generality when we have non-zero values of NSNS two form with components along 
null direction. For that
reason we restrict ourselves to the case of zero NSNS two form along these directions 
leaving the analysis of the most general case to the future. However even in the case of zero NSNS two form we derive interesting results. Explicitly, we show that T-dual string is either
relativistic or non-relativistic in accord with the  form of the background fields. In more details,  in case of the background \cite{Harmark:2019upf} we find that T-dual string
has the same form as the original one when component of the clock form $\tau_\mu dx^\mu$ along 
spatial direction $y$, where we perform T-duality, is zero. This result is in agreement with the condition 
that was derived independently in \cite{Kluson:2019avy} when T-duality of effective action for non-relativistic D-branes was studied. In the opposite case we obtain ordinary relativistic string in the modified background whose explicit transformation 
rules are determined by solving equations of motion for two auxiliary fields. 

Let us outline our results and suggest possible extension of this work. First of all we applied  general analysis that had been studied in \cite{Kluson:2019qgj} to the case of the null background \cite{Harmark:2019upf} and we derived Lagrangian for non-relativistic string
in torsional NC geometry. Then we studied properties of this string under T-duality along spatial direction with isometry and we argued that T-dual string is either relativistic or non-relativistic with dependence on the value of the background clock form $\tau_\mu dx^\mu$. In case when the T-dual string is again non-relativistic string in torsional NC background we found T-dual background fields whose transformation rules are in agreement with Buscher's rules \cite{Buscher:1987sk,Buscher:1987qj} and also with 
the analysis performed in \cite{Kluson:2019qgj} which is nice consistency check of both approaches. On the other hand we should stress that these results were derived on condition when the NSNS two form with components along null and spatial directions are zero. We leave an analysis of the most general case to the future work. 

This paper is organized as follows. In the next section (\ref{second}) we
show how Lagrangian for non-relativistic string in torsional NC background can 
be derived using T-duality transformations along null directions. Then in section 
(\ref{third}) we study T-duality of this string along spatial direction and 
we determine T-dual components of the background fields. 
\section{String with Light-like Isometry and Non-Relativistic String}\label{second}
We begin with the bosonic string in the background with null isometry 
whose dynamics is governed by the Lagrangian density 
\begin{eqnarray}\label{Lnull}
& &\mL=-\frac{T}{2}N\sqrt{\omega}
[-\nabla_n x^\mu G_{\mu\nu}\nabla_n x^\nu+\frac{1}{\omega}
\partial_\sigma x^\mu\partial_\sigma x^\nu G_{\mu\nu}-2\nabla_n x^\mu
G_{\mu u}\nabla_n u+\frac{2}{\omega}\partial_\sigma x^\mu G_{\mu u}\partial_\sigma u]-
\nonumber \\
& & -TB_{\mu\nu}\partial_\tau x^\mu \partial_\sigma x^\nu-TB_{\mu u}\partial_\tau x^\mu\partial_\sigma u-TB_{u\mu}\partial_\tau u\partial_\sigma x^\mu  \ ,   \nonumber \\
\end{eqnarray}
where $T$ is string tension, $G_{\mu\nu} \ , B_{\mu\nu} \ , \mu,\nu=0,\dots,d-1$ are background metric and NSNS two form. Further, $G_{\mu u}$ and $B_{\mu u}$ are
background fields along null direction labelled with $u$. Finally we used 
$1+1$ decomposition of the world-sheet metric where $N$ is two dimensional lapse, 
$N^\sigma$ is two dimensional shift and $\omega$ is spatial component of the metric,
and $\nabla_n=\frac{1}{N}(\partial_\tau-N^\sigma \partial_\sigma)$, where $\tau$ and $\sigma$ label time and space dimensions on the string world-sheet, see 
\cite{Kluson:2019qgj} for more details.

The crucial point which is
related to this Lagrangian density is an absence of the metric component $G_{uu}$. This 
fact makes the Hamiltonian analysis of this theory rather problematic. Then it was  shown in \cite{Kluson:2019qgj}  that it is convenient to rewrite 
(\ref{Lnull}) into equivalent form 
\begin{eqnarray}\label{Lnullextended}
& &\mL=-\frac{T}{2}N\sqrt{\omega}
[-\nabla_n x^\mu \hG_{\mu\nu}\nabla_n x^\nu+\frac{1}{\omega}
\partial_\sigma x^\mu\partial_\sigma x^\nu\hG_{\mu\nu}-2\nabla_n x^\mu
\hG_{\mu u}\nabla_n u+\frac{2}{\omega}\partial_\sigma x^\mu\hG_{\mu u}\partial_\sigma u-\nonumber \\
& &-\nabla_n u \hG_{uu}\nabla_n u+\frac{1}{\omega}
\partial_\sigma u \hG_{uu}\partial_\sigma u+
\lambda^+ \bA+\lambda^-\bB+\lambda^+\lambda^-]-\nonumber \\
& &-T\hat{B}_{\mu\nu}
\nabla_n x^\mu \partial_\sigma x^\nu-T\hat{B}_{\mu u}\nabla_n x^\mu\partial_\sigma u-
T\hat{B}_{u\nu}\nabla_n u\partial_\sigma x^\nu \ , 
\nonumber \\
\end{eqnarray}
where now we have to choose $\bA$ and $\bB$ in such a way to ensure
that $\hG_{uu}=0$ after solving equations of motion for $\lambda^+$ and
$\lambda^-$. Further, we also demand that $\hG_{\mu\nu},\hB_{\mu\nu},\hG_{\mu u},
\hB_{\mu u}$ reduce to $G_{\mu\nu},B_{\mu\nu},G_{\mu u},B_{\mu u}$ when 
the equations of motion for $\lambda^+$ and $\lambda^-$ are solved.  More explicitly,
the equations of motion for $\lambda^+,\lambda^-$ have the form 
\begin{equation}
\lambda^-=-\bA \ , \quad \lambda^+=-\bB
\end{equation}
and hence they give following contribution to the Lagrangian density 
\begin{equation}
\lambda^+\bA+\lambda^-\bB+\lambda^+\lambda^-=-\bA\bB \ . 
\end{equation}
Generally $\bA$ and $\bB$ could have the form 
\begin{eqnarray}
\bA=\nabla_n x^\mu \bA_\mu+\nabla_n u \bY^+-\frac{1}{\sqrt{\omega}}
[\partial_\sigma x^\mu\bA_\mu+\partial_\sigma y \bY^+] \ , 
\nonumber \\
\bB=\nabla_n x^\mu \bB_\mu+\nabla_n u \bY^-+\frac{1}{\sqrt{\omega}}
[\partial_\sigma x^\mu\bB_\mu+\partial_\sigma u \bY^-] \ .
\nonumber \\
\end{eqnarray}
However as was shown in \cite{Kluson:2019qgj} it is sufficient 
to consider the case when $\bA_\mu=\bB_\mu=0$ since their non-zero
values modify the background metric $G_{\mu\nu}$ only. 
%
In this case we obtain following contribution to the Lagrangian 
after solving equations of motion for $\lambda^+,\lambda^-$ in the form 
\begin{eqnarray}
-\bA\bB=-[\nabla_n u \nabla_n u-\frac{1}{\omega}\partial_\sigma u
\partial_\sigma u]\bY^+\bY^-
\end{eqnarray}
which implies that the
Lagrangian density (\ref{Lnullextended}) reduces to the original one when 
\begin{equation}
\bY^+=\sqrt{\hG_{uu}} \ , \quad  \bY^-=-\sqrt{\hG_{uu}}.
\end{equation}
Note also that in this case all hatted and unhatted components of the background fields
coincide.   In other words, an extended action has the form 
\begin{eqnarray}\label{Lextenmin}
& &\mL=-\frac{T}{2}N\sqrt{\omega}
[-\nabla_n x^\mu \hG_{\mu\nu}\nabla_n x^\nu+\frac{1}{\omega}
\partial_\sigma x^\mu\partial_\sigma x^\nu\hG_{\mu\nu}-2\nabla_n x^\mu
\hG_{\mu u}\nabla_n u+\frac{2}{\omega}\partial_\sigma x^\mu\hG_{\mu u}\partial_\sigma u-\nonumber \\
& &-\nabla_n u \hG_{uu}\nabla_n u+\frac{1}{\omega}
\partial_\sigma u \hG_{uu}\partial_\sigma u+\nonumber \\
&&+\lambda^+(\nabla_n u-\frac{1}{\sqrt{\omega}}\partial_\sigma u)\bY^++\lambda^-
(\nabla_n u+\frac{1}{\sqrt{\omega}}\partial_\sigma u)\bY^-
+\lambda^+\lambda^-]-\nonumber \\
& &-T\hat{B}_{\mu\nu}
\nabla_n x^\mu \partial_\sigma x^\nu-T\nabla_n x^\mu B_{\mu u}\partial_\sigma u-
\nabla_n u B_{u\nu}\partial_\sigma x^\nu \ .\nonumber \\
\nonumber \\
\end{eqnarray}
This action is the starting point of our analysis which is based on the canonical
description of T-duality. The Hamiltonian corresponding to the action 
(\ref{Lextenmin}) was found in \cite{Kluson:2019qgj} and has the form 
\begin{eqnarray}
& &H=\int d\sigma (N^\tau \mH_\tau+N^\sigma \mH_\sigma) \ , 
\nonumber \\
& &\mH_\tau=
 \pi_\mu \hG^{\mu\nu}\pi_\nu+2\pi_\mu \hG^{\mu u}
(\pi_u+\frac{T}{2}\tlambda^+\bY^++
\frac{T}{2}\tlambda^-\bY^-)+\nonumber \\
& &+(\pi_u+\frac{T}{2}\tlambda^+\bY^++
\frac{T}{2}\tlambda^-\bY^-)\hG^{uu}
(\pi_u+\frac{T}{2}\tlambda^+\bY^++
\frac{T}{2}\tlambda^-\bY^-)+\nonumber \\
& &+T^2\partial_\sigma x^\mu\hG_{\mu\nu}\partial_\sigma x^\nu
+2T^2\partial_\sigma x^\mu \hG_{\mu u}\partial_\sigma u+
T^2\partial_\sigma u \hG_{uu}\partial_\sigma u-\nonumber \\
& &-T^2\tlambda^+\partial_\sigma u \bY^++
T^2\tlambda^-\partial_\sigma u \bY^-+T^2\tlambda^+\tlambda^-  \ , \quad 
 \mH_\sigma=p_\mu\partial_\sigma x^\mu+p_u\partial_\sigma u \ ,   
\nonumber \\
\end{eqnarray}
where we performed rescaling 
\begin{equation}
\sqrt{\omega }\lambda^+=\tlambda^+ \ , \quad 
\sqrt{\omega}\lambda^-=\tlambda^-
\end{equation}
and where $p_\mu$ and $p_u$ are momenta conjugate to $x^\mu,u$ respectively and where
\begin{equation}
\pi_\mu=p_\mu+T\hB_{\mu\nu}\partial_\sigma x^\nu+T\hB_{\mu u}\partial_\sigma u \ ,
\quad 
\pi_u=p_u+T\hB_{u\mu}\partial_\sigma x^\mu \ . 
\end{equation}

Having identified canonical Hamiltonian we can proceed to the definition 
of non-relativistic string when we perform T-duality along $u$ direction. 
This is done when we introduce dual coordinate $\eta$ and $p_\eta$ that are
related to $p_u$ and $u$ by canonical transformations
\cite{Alvarez:1994wj,Alvarez:1994dn}
\begin{equation}
p_u=-T\partial_\sigma \eta \ , \quad p_{\eta}=-T\partial_\sigma u \ . 
\end{equation}
Then performing again inverse Legendre transformation 
to Lagrangian description we obtain T-dual Lagrangian in the form 
\begin{eqnarray}
& &\mL'
=\frac{1}{4N^\tau}(g'_{\tau\tau}-2N^\sigma g'_{\tau\sigma}+(N^\sigma)^2
g'_{\sigma\sigma})-N^\tau T^2 g'_{\sigma\sigma}-T \hB'_{MN}\partial_{\tau}\tx^M
\partial_\sigma \tx^N
 +\nonumber \\
& &+\frac{T}{2}N\tlambda^+(\nabla_n\tx^M \bA_M-2T\partial_\sigma \tx^M \bA_M)
+\frac{T}{2}N\tlambda^-(\nabla_n\tx^M \bB_M+2T\partial_\sigma \tx^M\bB_M) \ , 
\nonumber \\
\end{eqnarray}
where 
\begin{equation}
g'_{\alpha\beta}=\hG'_{MN}\partial_\alpha \tx^M\partial_\beta \tx^N \ , \quad 
\tx^M\equiv (x^\mu,\eta)
\end{equation}
and where the background fields are given by standard Buscher's rules
\cite{Buscher:1987sk,Buscher:1987qj}
\begin{eqnarray}
& &\hG'_{\mu\nu}=\hG_{\mu\nu}-\frac{1}{\hG_{uu}}\hG_{\mu u}\hG_{u \nu}+
\frac{1}{\hG_{uu}}\hat{B}_{\mu u}\hat{B}_{\nu u} \ , \nonumber \\
& & 
\hG'_{\mu \eta}=\frac{\hat{B}_{\mu u}}{\hG_{uu}} \ ,  \quad 
\hG'_{\eta\mu}=-\frac{\hat{B}_{u 
\nu}}{\hG_{uu}} \ , \quad 
\hG'_{\eta\eta}=\frac{1}{\hG_{uu}} \nonumber \\
& & \hB'_{\mu\nu}=\hB_{\mu\nu}-\frac{\hG_{\mu u}}{\hG_{uu}}\hat{B}_{u \nu}-\frac{\hB_{\mu u}}{\hG_{uu}}\hG_{u\nu } \ , 
\nonumber \\
& & \hB'_{ \mu \eta}=\frac{\hG_{\mu\eta}}{\hG_{uu}} \ , \quad \hB'_{\eta \nu}=-\frac{\hG_{u \nu}}{\hG_{uu}} \ . 
\nonumber \\
\end{eqnarray}
Finally, $\bA_M$ and $\bB_M$ are defined as 
\begin{equation}\label{defbA}
\bA_M=\left(\frac{1}{\sqrt{\hG_{uu}}}(\hG_{\mu u}-\hat{B}_{\mu u}),-\frac{1}{\sqrt{\hG_{uu}}}\right) \ , 
\quad 
\bB_M=-\left(\frac{1}{\sqrt{\hG_{uu}}}(\hG_{\mu u}+\hat{B}_{\mu u}),\frac{1}{\sqrt{\hG_{uu}}}\right) \ .
\end{equation}
Let us now solve the equations of motion for 
 $\tlambda^+,\tlambda^-$ that have the form
\begin{equation}\label{eqlambda}
\nabla_n \tx^M \bA_M=2T\partial_\sigma \tx^M \bA_M \ , \quad 
\nabla_n \tx^M \bB_M=-2T\partial_\sigma \tx^M \bB_M \ . 
\end{equation}
If we multiply the first equation with $\partial_\sigma \tx^M\bB_M$ and the second one by $\partial_\sigma \tx^N\bA_N$ and sum them we obtain 
\begin{eqnarray}
N^\sigma=\frac{\bM_{\tau\sigma}}{\bM_{\sigma\sigma}}  \ , 
\quad 
\bM_{\alpha\beta}=\partial_\alpha \tx^M \bM_{MN}\partial_\beta \tx^N \ , \\ \nonumber \\
\end{eqnarray}
where the matrix $\bM_{MN}$ is defined as 
\begin{equation}
\bM_{MN}=\frac{1}{2}(\bA_M\bB_N+\bB_M\bA_N)  \ .
\nonumber \\
\end{equation}
Further,if we multiply two equations in (\ref{eqlambda}) we obtain 
\begin{equation}
N^\tau=\frac{\sqrt{-\det \bM_{\alpha\beta}}}{2\bM_{\sigma\sigma}} \ . 
\end{equation}
Then the final  Lagrangian density has the form of non-relativistic string action 
\begin{eqnarray}
\mL^T=-\frac{T}{2}\sqrt{-\det\bM}\bM^{\alpha\beta}g'_{\alpha\beta}-T\hB' _{\mu\nu}
\partial_\tau \tx^\mu\partial_\sigma \tx^\nu-T \hB'_{\mu\eta}\partial_\tau x^\mu
\partial_\sigma \eta-T\hB'_{\eta\mu}\partial_\tau \eta\partial_\sigma x^\mu , 
\nonumber \\
\end{eqnarray}
where $\bM^{\alpha\beta}$ is matrix inverse to $\bM_{\alpha\beta}$ so that
$\bM^{\alpha\beta}\bM_{\beta\gamma}=\delta^\alpha_\beta$.

We use this general procedure for the case of the background with null isometry which 
defines NC with torsion  
 \cite{Harmark:2017rpg,Harmark:2019upf}
\begin{equation}\label{backfield}
ds^2=g_{MN}dx^M dx^N=2\tau (du-m)+h_{\mu\nu}dx^\mu dx^\nu\ ,  \quad 
\tau=\tau_\mu dx^\mu \ , \quad  m=m_\mu dx^\mu \ , 
\end{equation}
where $\det h_{\mu\nu}=0$. 
We also have non-zero NSNS two form with following components
\begin{equation}
\hB_{\mu\nu} \ , \quad \hB_{u\mu}=b_\mu \ . 
\end{equation}
For this background the components of the matrix $\bM_{MN}$ have the form 
\begin{equation}
\bM_{\mu\nu}
=-\frac{1}{\hG_{uu}}
(\tau_\mu\tau_\nu-b_\mu b_\nu) \ , \quad 
\bM_{\mu \eta}=-\frac{b_\mu}{\hG_{uu}} \ , \quad \bM_{\eta\eta}=\frac{1}{\hG_{uu}}  \ . 
\end{equation}
 Without lost of generality we can take $\hG_{uu}=1$ and hence we obtain 
 \begin{eqnarray}
 \bM_{\alpha\beta}=(b_\alpha-\partial_\alpha \eta)(b_\beta-\partial_\beta \eta)-
 \tau_\alpha\tau_\beta \ , \nonumber \\
    \end{eqnarray}
%
so that 
  \begin{eqnarray}
  \det \bM=-((\partial_\tau \eta-b_\tau)\tau_\sigma-
  \tau_\tau(\partial_\sigma \eta-b_\sigma))^2 \ . 
  \nonumber \\
  \end{eqnarray}
  Further, with the help of the background fields (\ref{backfield}) 
  we obtain  
  \begin{eqnarray}
& &  \hg'_{\alpha\beta}=\hh_{\alpha\beta}-\tau_\alpha\tau_\beta+(\partial_\alpha \eta-b_\alpha)
  (\partial_\beta \eta-b_\beta) \ , 
  \nonumber \\
 & & \hB'_{\mu\nu}=\hB_{\mu\nu}-\tau_\mu b_\nu+\tau_\nu b_\mu \ , \nonumber \\
& &  \hB'_{\mu\nu}\partial_\tau x^\mu\partial_\sigma x^\nu+\hB'_{\mu \eta}\partial_\tau x^\mu
  \partial_\sigma \eta+\hB'_{\eta\mu}\partial_\tau \eta\partial_\sigma x^\mu=
  \nonumber \\
& &=\hB_{\tau\sigma}+\tau_\sigma(b_\tau-\partial_\tau \eta)-\tau_\tau(b_\sigma-
  \partial_\sigma \eta) \ ,  \nonumber \\
  \end{eqnarray}
  where $\hh_{\alpha\beta}=\hh_{\mu\nu}
  \partial_\alpha x^\mu\partial_\beta x^\nu \ , \hh_{\mu\nu}=h_{\mu\nu}-\tau_\mu m_\nu-\tau_\nu m_\mu$. Then we see that
   Lagrangian density is equal to
  \begin{eqnarray}\label{mLnontor}
& &\mL
=\frac{T}{2
	((\partial_\tau \eta-b_\tau)\tau_\sigma-\tau_\tau
	(\partial_\sigma \eta-b_\sigma	))}\times \nonumber \\
& & [((\partial_\sigma\eta-b_\sigma)(\partial_\sigma \eta-b_\sigma)-\tau_\sigma\tau_\sigma)
\hh_{\tau\tau}-2((\partial_\sigma \eta-b_\sigma)(\partial_\tau \eta-b_
\tau)-\tau_\sigma\tau_\tau)\hh_{\tau\sigma}+\nonumber \\
& &((\partial_\tau \eta-b_\tau)
(\partial_\tau \eta-b_\tau)-\tau_\tau\tau_\tau)\hh_{\sigma\sigma}] 
\nonumber \\
& &=\frac{T}{2
 	((\partial_\tau \eta-b_\tau)\tau_\sigma-\tau_\tau
 	(\partial_\sigma \eta-b_\sigma	)}\times
 \epsilon^{\alpha\alpha'}\epsilon^{\beta\beta'}
 ((\partial_\alpha \eta-b_\alpha)(\partial_\beta \eta-b_\beta)-\tau_\alpha
 \tau_\beta)\hh_{\alpha'\beta'} \  \ ,  \nonumber \\
\end{eqnarray}  
were $\epsilon^{\alpha\beta}=-\epsilon^{\beta\alpha} \ , \epsilon^{01}=1$. We see
that the Lagrangian density (\ref{mLnontor}) agrees with the Lagrangian density found in \cite{Harmark:2019upf} and
we mean that this is nice consistency check of the general procedure outlined above. In the next section we analyse
how this non-relativistic string transforms under T-duality.
\section{T-duality of Non-Relativistic String in Torsional NC Geometry}\label{third}
We see that the Lagrangian density given above is non-linear and hence
it is not clear how to use standard procedure based on the gauging of the isometry direction. For that reason we mean that it is natural to start with extended
relativistic string and perform T-duality along both two directions, one corresponding to the original dimension that defines non-relativistic string and the
second one that corresponds to T-duality along spatial dimension. Since we study
this problem with the help of the canonical formalism 
we start with the Hamiltonian for extended string and perform T-duality along two directions, one corresponding to $u$ and the second one to spatial direction that we label as $y$. 
We use common notation where 
\begin{equation}
\tp^m=(p_{\ty},p_\eta) \ , \quad 
\tx_m=(\ty,\eta) \ ,  m=\ty,\eta \ .
\end{equation}
Finally we define  $\bY^\pm_m$ as $\bY^\pm_m=(0,\bY^\pm)$. 
We also restrict ourselves to the case when $\hB_{\mu u}=0$.
Then T-dual Hamiltonian
constraint has the form 
\begin{eqnarray}\label{mHTdual}
& &\mH_\tau^{T}=\mH_\tau(p_y=-T\partial_\sigma \ty,p_u=-T\partial_\sigma \eta,
\partial_\sigma y=-T^{-1}p_{\ty}, \partial_\sigma u=-T^{-1}p_\eta)=
\nonumber \\
& &=(k'_i-\hB_{im}\tp^m)\hG^{ij}
(k'_j-\hB_{jn}\tp^n)
+\nonumber \\
& &-2T(k'_i-\hB_{im}\tp^m)\hG^{in}\bW_n
+T^2\bW_m \hG^{mn}\bW_n+\nonumber \\
& &+T(k_i'-\hB_{in}\tp^n)\hG^{im}(\tlambda^+\bY^+_m+
\tlambda^-\bY^-_m)-
T^2\bW_m\hG^{mn}(\tlambda^+\bY^+_n+\tlambda^-\bY^-_n)+\nonumber \\
& &+\frac{T^2}{4}(\tlambda^+\bY^+_m+\tlambda^-\bY^-_m)
\hG^{mn}(\tlambda^+\bY^+_n+\tlambda^-\bY^-_n)+
T\tlambda^+\tp^m \bY^+_m-T\tlambda^-
\tp^m \bY^-_m+T^2\tlambda^+\tlambda^-+\nonumber \\
& &+T^2\partial_\sigma x^i \hG_{ij}\partial_\sigma x^j-2T
\partial_\sigma x^i\hG_{im}\tp^m+
\tp^m\hG_{mn}\tp^{n} \ , \nonumber \\
& &\mH_\sigma^T=p_i \partial_\sigma x^i+p_{\ty}\partial_\sigma \ty+
p_\eta \partial_\sigma \eta \ , \nonumber \\
\end{eqnarray}
where $k'_i=p_i+T\hB_{ij}\partial_\sigma x^j, i,j,k=0,\dots,d-2 $
 and where 
\begin{equation}
\bW_m=\partial_\sigma \tx_m-B_{mi}\partial_\sigma x^i \ .
\end{equation}
In order to determine
form of the background fields it is convenient to derive corresponding Lagrangian
from (\ref{mHTdual}). We begin 
with the equations of motion for $\tx_m,x^i$
\begin{eqnarray}\label{eqxm}
& &\partial_\tau x^i=\pb{x^i,H^T}=
2N^\tau \hG^{ij}(k'_j-\hB_{jm}\tp^m)-2TN^\tau \hG^{im}\bW_m
+\nonumber \\
& &+TN^\tau\hG^{im}(\tlambda^+\bY_m^++\tlambda^-\bY^-_m)+N^\sigma \partial_\sigma x^i  \ , \nonumber \\
& &\partial_\tau \tx_m=\pb{\tx_m,H^T}=2N^\tau\hB_{mi}\hG^{ij}
(k_j-\hB_{jn}\tp^n)-
2TN^\tau \hB_{mi}\hG^{in}\bW_n+\nonumber \\
&&+
TN^\tau \hB_{mi}\hG^{in}(\tlambda^+\bY^+_n+\tlambda^-\bY_n^-)+TN^\tau (\tlambda^+ \bY^+_m-\tlambda^-\bY^-_m) \nonumber \\
& &-2TN^\tau \hG_{mi}\partial_\sigma x^i+
2N^\tau \hG_{mn}\tp^n+N^\sigma \partial_\sigma \tx_m \ ,
\nonumber \\
\end{eqnarray}
where $H^T=\int d\sigma (N^\tau \mH_\tau^T+N^\sigma \mH_\sigma^T)$.
If we combine (\ref{eqxm}) together we can express 
 $\tp^m$ as 
\begin{eqnarray}
\tp^m=\frac{1}{2N^\tau}\tG^{mn}
(\tX_n-\hB_{ni}X^i+2TN^\tau \hG_{ni}\partial_\sigma x^i-TN^\tau (\tlambda^+
\bY_n^+-\tlambda^-\bY_n^-)) \ , \nonumber \\
\end{eqnarray}
where $\tX_n=\partial_\tau \tx_n-N^\sigma\partial_\sigma \tx_n \ , 
  X^i=\partial_\tau x^i-N^\sigma \partial_\sigma x^i$ and 
where we introduced matrix inverse $\tG^{mn}$ to $\hG_{mn}$
\begin{equation}
\tG^{mn}\hG_{np}=\delta^m_p \ . 
\end{equation}
We further introduce matrix $\tG_{ij}$ inverse to $\hG^{ij}$ 
\begin{equation}
\tG_{ij}=\hG_{ij}-\hG_{im}\tG^{mn}\hG_{mj} \ , \quad 
\tG_{ij}\hG^{jk}=\delta_i^k \ , \quad  \tG_{ij}\hG^{jm}=-\hG_{in}\tG^{nm}
\end{equation}
so that
\begin{eqnarray}
k'_i-\hB_{im}\tp^m=\frac{1}{2N^\tau}
\tG_{ij}
(X^j+2TN^\tau \hG^{jm}\bW_m-TN^\tau\hG^{jm}
(\tlambda^+\bY_m^++\tlambda^-\bY_m^-)) \ . 
\nonumber \\
\end{eqnarray}
Then, after some algebra, we find Lagrangian density in the form 
\begin{eqnarray}\label{mLTdual}
& &\mL^T=\tp^m\partial_\tau \tx_m+p_i\partial_\tau x^i-N^\tau\mH^T_\tau-N^\sigma \mH^T_\sigma=
\nonumber \\
& &=\frac{1}{4N^\tau}(X^i\hG'_{ij}X^j+\tX_m\hG'^{mn}\tX_n+\tX_m \hG'^m_{ \ i}X^i+
X^i\hG'^{ \ n}_i\tX_n)-\nonumber \\
& &N^\tau T^2(\partial_\sigma \tx_m \hG'^{mn}\partial_\sigma \tx_n+
\partial_\sigma \tx_m \hG'^m_{ \ i}\partial_\sigma x^i+
\partial_\sigma x^i\hG'^{ \ m}_i\partial_\sigma \tx_m
+\partial_\sigma x^i\hG'_{ij}\partial_\sigma x^j)-
\nonumber \\
& &-T\hB'_{ij}\partial_\tau x^i\partial_\sigma x^j-T\hB'^{ \ m}_i\partial_\tau x^i
\partial_\sigma \tx_m-T\hB'^m_{ \ i}\partial_\tau \tx_m\partial_\sigma x^i+
\nonumber \\
& &-\frac{T}{2}N^\tau\tlambda^+
(\nabla_n x^i(-\tG_{im}+\hB_{im})\hG^{mn}\bY_n^++\nabla_n \tx_n\tG^{nm}\bY_m^+ -\nonumber \\
& & -2\partial_\sigma\tx_m \tG^{mn}\bY_n^+
+2\partial_\sigma
x^i(\hG_{in}-\hB_{in})\tG^{mn}\bY_n^+)
\nonumber \\
& &-\frac{T}{2}N^\tau \tlambda^-
(\nabla_n x^i(-\tG_{im}-\hB_{im})\tG^{mn}\bY_n^--\nabla_n \tx_n\tG^{nm}\bY_m^-
-\nonumber \\
& &-2\partial_\sigma \tx_n\tG^{nm}\bY_m^--2\partial_\sigma x^i(\hG_{im}+\hB_{im})\tG^{mn}\bY_n^-)- \nonumber \\
& & -T^2N^\tau\tlambda^+\tlambda^-(\bY_m^+\tG^{mn}\bY_n^-+1) \ , 
\nonumber \\
\end{eqnarray}
where we have following components of background metric and NSNS two form
\begin{eqnarray}\label{Tdual}
& &\hG'_{ij}=\tG_{ij}-\hB_{in}\tG^{nm}\hB_{mj}=
\hG_{ij}-\hG_{in}\tG^{nm}\hG_{nj}-\hB_{in}\tG^{nm}
\hB_{mj} \ , \nonumber \\
& & \hG'^{mn}=\tG^{mn} \ ,  \quad 
\hG'^m_{ \ i}=-\tG^{mn}\hB_{ni} \ , \quad 
\hG'^{ \ n}_i=\hB_{im}\tG^{mn} \ , \nonumber \\
& &\hB'^{ \ m}_i=\hG_{in}\tG^{nm} \ , \quad \hB'^m_{ \ i}=-\hG^{mn}\hG_{mi} \ , 
\quad 
\hB'_{ij}=\hB_{ij}-\hB_{in}\tG^{nm}\hG_{mj}+\hG_{im}
\tG^{mn}\hB_{nj} \ . \nonumber \\
\end{eqnarray}
Now the nature of T-dual string depends on the form of the
inverse matrix $\tG^{mn}$. In case when $\tG^{\eta\eta}=\frac{1}{\hG_{uu}}$
we obtain that T-dual string is  non-relativistic string. To see this
we introduce again notation
\begin{eqnarray}
& &\tx^M=(x^i,\tx_m) \  , \quad \bA_M=((-\hG_{in}+\hB_{in})\tG^{nm}\bY_m^+,
\tG^{nm}\bY_m^+) \ ,  \nonumber \\
& &\bB_M=((-\hG_{in}-\hB_{in})\tG^{nm}\bY_m^-,-\tG^{nm}\bY_m^-) \ , \nonumber \\
\end{eqnarray}
where of course we could express $\bA_M$ and $\bB_M$ with the help of the transformed
fields given in (\ref{Tdual}). Then the  expression proportional to $\tlambda^+,\tlambda^-$ can be written as
\begin{eqnarray}\label{lambdacon}
& &-\frac{T}{2}N^\tau \tlambda^+(\nabla_n \tx^M \bA_M -2T\partial_\sigma \tx^M \bA_M)
-\nonumber \\
& &-\frac{T}{2}N^\tau \tlambda^+(\nabla_n \tx^M \bB_M+2T\partial_\sigma \tx^M \bB_M)-
T^2N^\tau \tlambda^+\tlambda^-(\bY_m^+\tG^{mn}\bY_n^-+1) \ . 
\nonumber \\
\end{eqnarray}
Now in the first case when $\tG^{\eta\eta}=\frac{1}{\hG_{uu}}$ we find
that $\bY_m^+\tG^{mn}\bY_n^-+1=0$ and hence 
we find
\begin{equation}
N^\sigma=\frac{\bM_{\tau\sigma}}{\bM_{\sigma\sigma}} \ , \quad 
N=\frac{\sqrt{-\det \bM_{\alpha\beta}}}{2\bM_{\sigma\sigma}}
\ , 
\end{equation}
where
\begin{equation}
\bM_{\alpha\beta}=\partial_\alpha \tx^M \bM_{MN}
\partial_\beta \tx^N \ , \quad 
\bM_{MN}=\frac{1}{2}(\bA_M \bB_N+\bB_M \bA_N) \ . 
\end{equation}
As a result we obtain Lagrangian density in the form
\begin{equation}
\mL^T=-\frac{T}{2}\sqrt{-\det \bM}\bM^{\alpha\beta}\hG'_{MN}
\partial_\alpha \tx^M \partial_\beta \tx^N
-T\hB'_{MN}\partial_\tau \tx^M \partial_\sigma \tx^N \ .
\end{equation}
In opposite case let us denote $\bX=(\bY_m^+\tG^{mn}\bY_n^-+1)$. Then 
the equation of motion for $\tlambda^\pm$ can be solved as
\begin{eqnarray}
\tlambda^-=-\frac{1}{2T\bX}(\nabla_n \tx^M \bA_M-2T\partial_\sigma \tx^M 
\bA_M) \ , \quad 
\tlambda^+=-\frac{1}{2T\bX}(\nabla_n\tx^M \bB_M+2T\partial_\sigma
\tx^M \bB_M) \ . 
\nonumber \\
\end{eqnarray}
Inserting this result into (\ref{lambdacon})  we obtain following contribution to the
Lagrangian density (\ref{mLTdual}) 
\begin{equation}
\frac{1}{2\bX}N^\tau(\nabla_n\tx^M \bA_M-2T\partial_\sigma \tx^M \bA_M)
(\nabla_n \tx^N \bB_N+2T\partial_\sigma \tx^N \bB_N)
\end{equation}
and hence we  obtain relativistic form of the Lagrangian density
\begin{eqnarray}
\mL=\frac{1}{4N^\tau}(\hg''_{\tau\tau}-N^\sigma \hg''_{\tau\sigma}+
(N^\sigma)^2 \hg''_{\sigma\sigma})-N^\tau T^2 
\hg''_{\sigma\sigma}-T \hB''_{MN}\partial_\tau \tx^M
\partial_\sigma \tx^N \ , \nonumber \\
\end{eqnarray}
where
\begin{eqnarray}
\hG''_{MN}=\hG'_{MN}+\frac{2}{\bX}\bM_{MN} \ ,  \quad 
\hB''_{MN}=\hB'_{MN}+\frac{1}{\bX}(\bA_N\bB_M-\bA_M\bB_N) \ .
\nonumber \\
\end{eqnarray}
Now we return to the background (\ref{backfield}). In this explicit case the matrix 
$\hG_{mn}$ has components 
\begin{equation}
\hG_{mn}=\left(\begin{array}{cc}
h_{yy}-2\tau_y m_y & \tau_y \\
\tau_y & \hG_{uu} \\ \end{array}
\right)
\end{equation}
so that inverse metric has the form 
\begin{equation}
\tG^{mn}=
\frac{1}{\det \hG_{mn}}
\left(\begin{array}{cc}
\hG_{uu} & -\tau_y \\
-\tau_y & h_{yy}-2\tau_y m_y \\ \end{array}\right) \ , 
\end{equation}
where
\begin{equation}
\det\hG_{mn}=
(h_{yy}-2\tau_y m_y)\hG_{uu}-\tau_y\tau_y \ . 
 \end{equation}
As it is clear from the matrix above the condition to have $\hG^{\eta\eta}=\frac{1}{\hG_{uu}}$ we should demand that $\tau_y=0$. 
so that
\begin{equation}
\tG^{mn}=
\left(\begin{array}{cc}
\frac{1}{h_{yy}} & 0 \\
0 & \frac{1}{\hG_{uu}} \\ \end{array}\right) \ .
\end{equation}
In this case we obtain following components of the vectors $\bA_M$ and 
$\bB_M$
\begin{eqnarray}
\bA_M=\left(-\frac{\hG_{iu}}{\sqrt{\hG_{uu}}},0,\frac{1}{\sqrt{\hG_{yy}}}\right) \ ,
\quad  
\bB_M=\left(\frac{\hG_{iu}}{\sqrt{\hG_{uu}}},0,\frac{1}{\sqrt{\hG_{yy}}}\right)
\nonumber \\
\end{eqnarray}
and hence matrix $\bM_{MN}$ has the form 
\begin{eqnarray}
& & \bM_{ij}=-\frac{\tau_i\tau_j}{\hG_{uu}} \ , \quad 
\bM_{i\ty}=0 \ , \quad \bM_{i\eta}=
\tau_i \ , \nonumber \\
& &\bM_{\ty\ty}=\bM_{\ty\eta}=0 \ , \quad \bM_{\eta\eta}=\frac{1}{\hG_{uu}} \ . 
\nonumber \\
\end{eqnarray}
When we choose $\hG_{uu}=1$ we obtain that T-dual background fields have the form 
\begin{eqnarray}\label{backTnon}
& &\hG'_{ij}
=\hh_{ij}-\frac{\hh_{iy}\hh_{yj}}{h_{yy}}-\tau_i\tau_j
-\frac{\hB_{iy}\hB_{yj}}{h_{yy}}\equiv
\hh'_{ij}-\tau_i \tau_j
 \ , \nonumber \\
& &\hG'^{\ty}_{ \ i}=-\frac{1}{h_{yy}}\hB_{yi} \ , 
\hG'^{ \ \ty}_i=\frac{1}{h_{yy}}\hB_{iy} \ , 
\nonumber \\
& &\hG'^{\eta}_{ \ i}=-\hB_{ui}=0 \ , \quad \hG'^{ \ \eta}_i=\hB_{iu}=0  \ , \nonumber \\
& &\hG'^{\eta\eta}=1 \ , \quad \hG'^{\ty\ty}=\frac{1}{h_{yy}} \ , \nonumber \\
& &\hB'_{ij}=\hB_{ij}-\frac{\hB_{iy}\hG_{yj}}{h_{yy}}-
\hB_{iu}\tau_j+\tau_i \hB_{uj}+\frac{\hG_{iy}\hB_{yj}}{h_{yy}}=
\hB_{ij}-\frac{\hB_{iy}\hh_{yj}}{h_{yy}}+\frac{\hh_{iy}\hB_{yj}}{h_{yy}}
 \ , \nonumber \\
& &\hB'^{ \ \ty}_i=\frac{\hG_{iy}}{h_{yy}}=\frac{\hh_{iy}}{h_{yy}} \ , \quad
\hB'^{ \ \eta}_i=\tau_i \ .  \nonumber \\
\quad 
\end{eqnarray}
It is instructive to find explicit form of the T-dual Lagrangian. First of all we have
\begin{equation}
\bM_{\alpha\beta}=(\partial_\alpha \eta-\tau_\alpha)(\partial_\beta \eta-\tau_\beta) \ .
\end{equation}
 Then we obtain 
\begin{eqnarray}
& &g'_{\alpha\beta}=\hh_{\alpha\beta}'-\tau_\alpha\tau_\beta+\partial_\alpha \eta\partial_\beta \eta \ , \nonumber \\
& &\hB'_{MN}\partial_\tau \tx^M\partial_\sigma \tx^N=\hB'_{\tau\sigma}+\tau_\tau
\partial_\sigma \eta-\tau_\sigma\partial_\tau \eta
\nonumber \\
\end{eqnarray}
and hence we see that T-dual Lagrangian density has the same form as in 
(\ref{mLnontor}) (with zero $b_\mu$) 
with background metric and NSNS two form fields given in (\ref{backTnon}). This is again 
nice consistency check. Of course we should stress that we do not consider
the most general case when $\hB_{u\mu}=b_\mu\neq 0$. On the other hand we do not expect
that the presence of non-zero NSNS two form field qualitatively changes the
transformation rules presented in this paper however this problem should be investigated
further.

\end{document}